\documentclass [preprint,aps,showpacs]{revtex4}
\usepackage[final]{graphics}
\usepackage{amssymb}
\usepackage{amsfonts}
\usepackage{epsfig}
\usepackage{graphicx}

\begin{document}
\title{Diffusive epidemic process: theory and simulation}
\author{Daniel Souza Maia\footnote{dfmaia@ufmg.br}
and Ronald Dickman\footnote{dickman@fisica.ufmg.br}}
\address{Departamento de F\'{\i}sica, Instituto de Ci\^encias Exatas,\\
Universidade Federal de Minas Gerais \\
C. P. 702, 30123-970, Belo Horizonte, Minas Gerais - Brazil }

\date{\today}

\begin{abstract}
We study the continuous absorbing-state phase transition in the one-dimensional
diffusive epidemic process via mean-field theory and Monte Carlo simulation. In
this model, particles of two species (A and B) hop on a lattice and undergo
reactions B $\to$ A  and A + B $\to$ 2B; the total particle number is
conserved. We formulate the model as a continuous-time Markov process described
by a master equation. A phase transition between the (absorbing) B-free state
and an active state is observed as the parameters (reaction and diffusion
rates, and total particle density) are varied.  Mean-field theory reveals a
surprising, nonmonotonic dependence of the critical recovery rate on the
diffusion rate of B particles. A computational realization of the process that
is faithful to the transition rates defining the model is devised, allowing for
direct comparison with theory. Using the quasi-stationary simulation method we
determine the order parameter and the survival time in systems of up to 4000
sites. Due to strong finite-size effects, the results converge only for large
system sizes. We find no evidence for a discontinuous transition.  Our results
are consistent with the existence of three distinct universality classes,
depending on whether A particles diffusive more rapidly, less rapidly, or at
the same rate as B particles.

\end{abstract}

\pacs{05.10.-a, 02.50.Ga, 05.40.-a, 05.70.Ln}

\maketitle

\section{Introduction}

This work is devoted to the one-dimensional diffusive epidemic process (DEP)
\cite{kree}, a model system in which two kinds of particles, A and B, diffuse
on a lattice and undergo reactions B $\to$ A and A + B $\to$ 2B.  There is no
intrinsic limit on the number of particles that may be present at a given site;
the total number of particles is conserved. In the epidemic interpretation A
represents a healthy organism and B and infected one, with the reactions above
corresponding, respectively, to spontaneous recovery and transmission of
disease on contact.   Other interpretations are possible, for example, A
could represent a properly folded protein and B a misfolded one, etc.

The DEP is a nonequilibrium stochastic model exhibiting a phase transition to
an absorbing state \cite{marro,hinrichsen,lubeck04,odor04}. Such phase
transitions arise in many models of epidemics, population dynamics and
autocatalytic chemical reactions, and have attracted much interest in
nonequilibrium statistical mechanics, in efforts to characterize the associated
universality classes.  The simplest example is the contact process (CP), or its
discrete-time version, directed percolation (DP).  In the CP, each site of a
lattice is either vacant (0) or occupied by a particle ($X$).  Particles die
(i.e., the reaction $X \to 0$) at rate 1, independent of the configuration of
the rest of the system, and reproduce at rate $\lambda$ (reaction $X + 0 \to
2X$).  The offspring particle survives if and only if the site it is sent to
(selected at random from the neighbors of the reproducing particle) is vacant.
The CP is therefore a minimal model for birth-and-death processes with local
competition for space. The particle-free configuration is absorbing. It is
known that for reproduction rate $\lambda < \lambda_c$, the stationary density
of particles $\rho$ is zero, and that (in the infinite system size limit),
$\rho$ grows continuously from zero as $\lambda$ is increased beyond
$\lambda_c$. Thus $\rho$ serves as the order parameter for this phase
transition.

Critical scaling in the contact process and allied models has been studied
extensively, both theoretically and numerically.  A central conclusion deriving
from these studies is that critical behavior of the DP type is generic for
models exhibiting a continuous phase transition to an absorbing state, in the
absence of any additional symmetries or conserved quantities.  (Note that the
absorbing state may in some cases encompass more that one configuration, as is
the case in the pair contact process \cite{pcp}.  If these configurations are
not related by any symmetry the critical behavior is still expected to fall in
the DP universality class.)  Particularly relevant in the present context is
the fact that the {\it diffusive} CP (in which particles hop at finite rate)
also belongs to the DP class \cite{iwandcp}.

The DEP offers a more complicated scenario of scaling. Let $D_A$ ($D_B$) denote
the diffusion rate of A (B) particles. An especially interesting aspect of the
DEP is that the absorbing-state phase transition appears to belong to three
distinct classes, depending on whether $D_A < D_B$, $D_A = D_B$, or $D_A >
D_B$.  The renormalization group (RG) analysis of Oerding et al. \cite{oerding}
predicts a continuous phase transition in the first two cases.  But for $D_A >
D_B$ there is no fixed point, leading to the conjecture that the transition is
discontinuous in this case. These authors provide numerical evidence of such a
transition in two dimensions; numerical studies of the one-dimensional DEP,
however, show a continuous phase transition \cite{fulco}.

The diffusive epidemic process was initially studied via RG methods
\cite{kree,oerding,wijland}.   Numerical simulations for equal diffusion rates
were reported in Ref. \cite{freitas}, yielding results in disagreement with the
RG prediction $\nu_\perp = 2$ (see \cite{janssende,freitasr}).  Subsequently,
the simulations reported in \cite{fulco} appeared to resolve this point, but
suggested other departures from RG predictions in the one-dimensional case.

Given the disagreement between theory and simulation, it is of interest to
perform further analysis of the DEP.  In the present work we study a
one-dimensional version of the process, formulated in continuous time, in a
manner allowing for a simple formulation of the master equation. We study this
process (defined in detail in the following section), using one- and two-site
mean-field theory (Sec. III), and Monte Carlo simulation (Sec. IV) using the
quasi-stationary simulation method. Conclusions and open questions are
discussed in Sec. V.

\section{Model}

The DEP is defined on a lattice of $L^d$ sites.  A configuration is specified
by the set of variables $a_j$ and $b_j$, denoting the number of A and B
particles at each site $j$.  There are no intrinsic restrictions on the number
of particles at each site, making this process, in the language commonly
employed in the literature, a ``bosonic" model.  (To avoid confusion we note
that quantum statistics do not enter into the problem.)  The model is a
continuous-time Markov process characterized by four kinds of transitions:

\begin{itemize}
\item Hopping of A particles to a randomly chosen nearest neighbor (NN) site, at rate $D_A$.

\item Hopping of B particles to a randomly chosen NN site, at rate $D_B$.

\item Transformation of B particles to A particles, at rate $r$.

\item Transformation of A particles to B particles, in the presence of a B
particle at the same site, at a rate of $\lambda$ per A-B pair.
\end{itemize}

\noindent This means that a given site $j$ loses (via diffusion) an A particle
at rate $D_A a_j$ (similarly for loss of a B particle), undergoes the process B
$\to$ A at rate $r b_j$, and the process A + B $\to$ 2B at rate $\lambda a_j
b_j$.  Note that all transitions conserve the total particle number $N = \sum_j
(a_j + b_j)$.

The process involves a rather large set of parameters: $D_A$, $D_B$, $r$, $\lambda$,
and the particle density $\rho = N/L^d$.  Since one of the rates may be eliminated
through a suitable scaling of time, we set $\lambda = 1$ from here on.  This still
leaves four control parameters.  In the studies described below, we fix the diffusion
constants and study the behavior in the $r$-$\rho$ plane, or fix $D_A$ and $\rho$ and
treat $D_B$ and $r$ as the control parameters.

\section{Mean-field theory}

We begin by studying the model using dynamic mean-field theory (MFT)
\cite{marro}. Here we present the results for one- and two-site approximations
in one dimension, which follow from the master equation for the probability
distribution $P(a_1,b_1;...;a_L,b_L,t)$.  Consider the evolution of the
one-site marginal distribution $P(a,b)$:

\begin{eqnarray}
\nonumber \dot{P}(a,b) &=& \!\! D_A [(a+1)P(a+1,b) - aP(a,b)] + D_A
\sum_{a',b'} a' [P(a-1,b;a',b') - P(a,b;a',b')]
\\
\nonumber &+& D_B [(b+1)P(a,b+1) - bP(a,b)] + D_B \sum_{a',b'} b'
[P(a,b-1;a',b') - P(a,b;a',b')]
\\
&+& r [(b\!+\!1) P(a\!-\!1,b\!+\!1) - b P(a,b)] +
(b\!-\!1)(a\!+\!1)P(a\!+\!1,b\!-\!1) - ab P (a,b) \label{mft1}
\end{eqnarray}
Here $P(a,b;a',b')$ is the joint probability distribution for a pair of NN
sites.  Eq.~(\ref{mft1}) is the first in a hierarchy of equations for the
probability distributions of 1, 2,...,$n$,... sites.  At each level of the
hierarchy, the diffusion terms couple the $n$-site distribution to that for
$n+1$ sites. One-site MFT truncates this hierarchy at lowest order, via the
factorization $P(a,b;a',b') = P(a,b)P(a',b')$, leading to,

\begin{eqnarray}
\nonumber \dot{P}(a,b) &=& \!\! D_A [(a+1)P(a+1,b) - aP(a,b)] + D_A \rho_A
[P(a-1,b) - P(a,b)]
\\
\nonumber &+& D_B [(b+1)P(a,b+1) - bP(a,b)] + D_B \rho_B [P(a,b-1) - P(a,b)]
\\
&+& r [(b\!+\!1) P(a\!-\!1,b\!+\!1) - b P(a,b)] +
(a\!+\!1)(b\!-\!1)P(a\!+\!1,b\!-\!1) - ab P(a,b) \label{mft2}
\end{eqnarray}
where $\rho_A = \sum_{a,b} a P(a,b)$ is the density of A particles and
similarly for $\rho_B$.  An equation for $\rho_A$ is found by multiplying the
above equation by $a$ and summing over all values of $a$ and $b$, giving,
\begin{equation}
\dot{\rho}_A = r \rho_B - \langle ab \rangle \label{mft3}
\end{equation}
where $\langle a^m b^n \rangle \equiv \sum_{a,b} a^m b^n P(a,b) $.  (Thus
$\langle a \rangle = \rho_A$ and similarly for $\rho_B$.)  Note that, as
discussed in detail below, the cross-moment $\langle a b \rangle$ is in general
different from the simple product of A and B particle densities. If we
nevertheless set $\langle a b \rangle = \rho_A \rho_B$, we have
\begin{equation}
\dot{\rho}_A = r\rho_{B} - \rho_{A} \rho_{B} \label{mft4}
\end{equation}
With the constraint $ \rho_A + \rho_B = \rho$, constant, we then find
\begin{equation}
\dot{\rho}_B = (\rho-r)\rho_{B} - \rho_{B}^2 \label{mft5}
\end{equation}
showing that at this (lowest) level of approximation the order parameter
$\rho_B$ satisfies the Malthus-Verhulst equation with reproduction rate $\rho -
r$.  At this level, which may be called a ``rate equation", a continuous phase
transition occurs at $\rho = r$, independent of the diffusion rates; the
stationary density of $B$ particles follows $\overline{\rho}_B = \rho - r$.
While the more detailed mean-field approximations described below yield a more
reliable prediction for the phase boundary, we observe, in all cases, a
continuous phase transition, and a linear relation between $\overline{\rho}_B$
and $\rho - r_c$ in the vicinity of the critical point (i.e., the usual
mean-field critical exponent $\beta = 1$). We find no hint of a discontinuous
transition.

A somewhat better result is obtained integrating (numerically) the full set of
one-site MF equations, Eq. (\ref{mft2}). In numerical analysis, it is necessary
to truncate this set of equations at cutoff values $a_c$ and $b_c$. This is
justified since the probability distribution falls off exponentially for large
$a$ and/or $b$.  The cutoff leads to certain technical restrictions in the
numerical analysis.  Naturally, transitions of the form $a_c \to a_c + 1$ must
be excluded from consideration. Moreover, a transition of the form $a \to a-1$,
due to an A particle hopping away from the site of interest, must have its rate
multiplied by $1 - P_A(a_c)$, where $P_A(a) = \sum_b P(a,b)$ is the one-site
marginal distribution for the number of A particles. (Similar restrictions
apply to transitions involving B particles.)

Once we take the full one-site probability distribution into account, the
results for the phase boundary $\rho_c(r)$ depend on the diffusion rates. Two
factors enter into this dependence.  First, in the vicinity of the phase
transition, the reaction terms cause the marginal distribution for the number
of B particles to deviate significantly from a Poisson distribution, while
rapid diffusion tends to make the distribution more Poisson-like.  Second, the
reactions cause the variables $a$ and $b$ to be anti-correlated [that is,
cov($a,b$) = $\langle ab \rangle - \rho_A \rho_B < 0$], whereas rapid diffusion
tends to eliminate this correlation.

Figure \ref{mft1} shows the critical line $\rho_c (r)$ as predicted by the
one-site MFT for various combinations of $D_A$ and $D_B$. The higher the
diffusion rates, the more closely $\rho_c$ approaches the simple rate equation
result $\rho_c = r$. For finite diffusion rates $\rho_c$ is always greater than
$\rho$, due again to the anti-correlation of $a$ and $b$. The critical value
$\rho_c$ appears to be more sensitive to $D_B$ than $D_A$. In fact, for $D_B =
0.2$, the curves for $D_A = 0.2$ and $D_A = 1$ are virtually identical. (For
larger values of $D_B$, increasing $D_A$ does reduce the critical density.)

\begin{figure}[h]
\epsfysize=8cm \epsfxsize=10cm \centerline{ \epsfbox{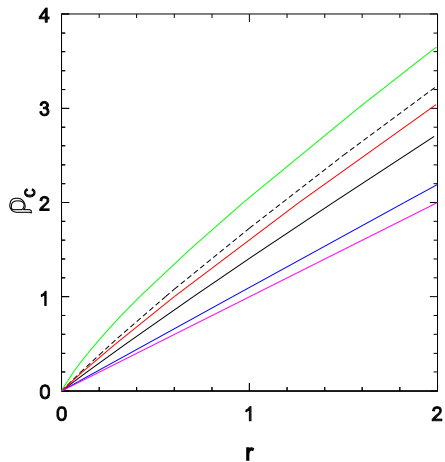}}
\caption{\footnotesize{Mean-field predictions for the critical particle density
$\rho_c$ versus recovery rate $r_c$.  Solid lines, top to bottom: $D_A = D_B =
0.2$; $D_A=0.2$, $D_B=1$; $D_A=1$, $D_B=1$; $D_A=D_B=5$; rate equation result,
$\rho_c = r$.  Bold dashed line: two-site mean-field theory for the case
$D_A=1$, $D_B=1$.}}
 \label{mft1}
\end{figure}

A richer description is obtained when we extend the approximation to two sites.
There are (in general) 16 transitions into (and out of) a given state
\protect{$(a,b;a',b')$}. Using the symmetry $P(a,b;a',b') =P(a',b';a,b)$, and
factoring three-site probabilities so that $P(a,b;a',b';a'',b'') =
P(a,b;a',b')P(a',b';a'',b'')/P(a',b')$, the equation governing the two-site
joint probability can be written as
\begin{eqnarray}
\frac{dP(a,b;a',b')}{dt}& = & \frac{D_A}{2}[(a+1)P(a+1,b;a',b') +(a'+1)P(a,b;a'+1,b') \nonumber \\
& & +(a'+1)P(a-1,b;a'+1,b') +(a+1)P(a+1,b;a'-1,b')] \nonumber \\
& & \frac{D_B}{2}[(b+1)P(a,b+1;a',b') +(b'+1)P(a,b;a',b'+1) \nonumber \\
& & +(b'+1)P(a,b-1;a',b'+1) +(b+1)P(a,b+1;a',b'-1)] \nonumber \\
& & + \frac{D_A}{2}\left[\Phi_A(a-1,b)P(a-1,b;a',b')
+ \Phi_A(a'-1,b')P(a,b;a'-1,b') \right] \nonumber \\
& & + \frac{D_B}{2}\left[\Phi_B(a,b-1)P(a,b-1;a',b')
+ \Phi_B(a',b'-1)P(a,b;a',b'-1) \right] \nonumber \\
& & + r[(b+1)P(a-1,b+1;a',b') + (b'+1)P(a,b;a'-1,b'+1)] \nonumber \\
& & + (a\!+\!1)(b\!-\!1)P(a\!+\!1,b\!-\!1;a',b') + (a'\!+\!1)(b'\!-\!1)P(a,b,a'\!+\!1,b'\!-\!1) \nonumber \\
& & - \left\{ D_A(a+a') + D_B (b+b') + \frac{D_A}{2} [\Phi_A(a,b) + \Phi_A(a',b')] \right. \nonumber \\
& & + \left. \frac{D_B}{2} [\Phi_B(a,b) + \Phi_B(a',b')] +  r(b+b') + ab + a'b'
\right\} P(a,b;a',b')
\end{eqnarray}

\noindent where
\begin{equation}
\Phi_A(a,b) = \frac{\sum_{a'}\sum_{b'} a'\,P(a,b,a',b')}{P(a,b)} \label{defphi}
\end{equation}

\noindent is the conditional A-particle density at a site, given that one of
its nearest neighbors has occupancy $(a,b)$.  ($\Phi_B$ is defined
analogously.) The above equations are integrated numerically using the
fourth-order Runge-Kutta method and a cutoff of 10 for the variables $a$, $b$,
$a'$ and $b'$. For densities $\rho \leq 2$, the error incurred is negligible.
Figure 1 shows that the two-site approximation predicts a larger value of
$\rho_c$ than does the one-site approximation, other parameters being equal.

We compare the mean-field predictions for the critical recovery rate against
simulation for three representative cases (all for density $\rho=1$) in Table
I. The site approximation overestimates $r_c$ by a factor of up to 3.3; in each
case the two-site approximation yields a substantial improvement, although it
still overestimates the critical value by a factor of up to 2.3.  In principle,
further improvement could be furnished using higher order approximations, but
in the present case the computational demands seem excessive.  [The number of
equations to be integrated in the $n$-site approximation is $[(a_c+1)(
b_c+1)]^n$.]

An important aspect of the model is the anti-correlation between variables $a$
and $b$ at a given site.  To quantify this we study
\begin{equation}
Q_A \equiv \frac{\langle a b \rangle}{\rho_B} - \rho_A =
\frac{\mbox{cov}(a,b)}{\rho_B} \label{defQA}
\end{equation}
At the critical point ($\rho_B \to 0$), $Q_A$ represents the excess density of
A particles at a site bearing a B particle.  In the one-site approximation, we
find (at the respective critical points), $Q_A$ = -0.583, -0.298, and -0.0895,
for $D_A\!=\!D_B$ = 0.2, 1, and 5, respectively.  That is, the species are
anti-correlated, and the magnitude of the correlation decreases with increasing
diffusion rate, as expected.  This effect appears even more markedly in the
two-site approximation, where, for example, $Q_A = -0.447$ at the critical
point with $D_A = D_B = 1$. (All results quoted are for density $\rho=1$.)
Similar values are found in simulations. The two-site approximation shows that
the anti-correlation of the numbers of A and B particles extends to the nearest
neighbor site: $\langle a_j b_{j+1}\rangle/\rho_B - \rho_A = -0.444$ for the
parameters noted above. The variables $b_j$ and $b_{j+1}$, on the other hand,
show a strong positive correlation.

\begin{figure}[h]
\epsfysize=8cm \epsfxsize=10cm \centerline{ \epsfbox{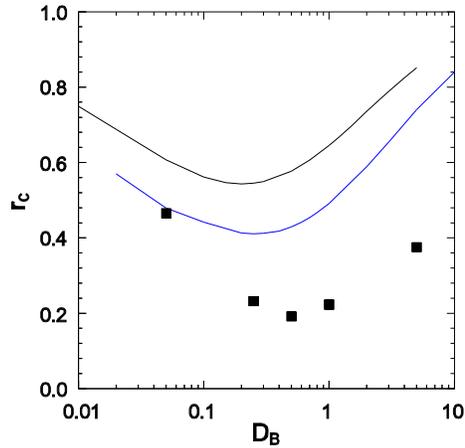}}
\caption{\footnotesize{Critical recovery rate $r_c$ versus B particle diffusion
rate $D_B$, for $D_A=0.5$ and $\rho=1$.  Upper curve: one-site MFT; lower
curve: two-site MFT; points: simulation.}}
 \label{epg}
\end{figure}

Since diffusion of B particles is essential to the survival of the process, one
might suppose that as $D_B$ is reduced, the critical density would increase
(for fixed recovery rate $r$), or the critical recovery rate $r_c$ decrease
(for fixed density). Mean-field theory (at both levels) predicts otherwise, as
shown in Fig.~\ref{epg}: the critical recovery rate exhibits a minimum at a
small value, $D_B^*$, of the B diffusion rate, but for even smaller values, it
grows rapidly. For $D_B > D^*$ the critical recovery rate grows systematically,
and appears to saturate at the rate-equation value, $r_c=\rho$. Fig. \ref{epg}
shows that $r_c$ follows the same qualitative trends in simulations as in the
mean-field approximations, although its numerical value is (generally)
considerably smaller.

\begin{figure}[h]
\epsfysize=8cm \epsfxsize=10cm \centerline{ \epsfbox{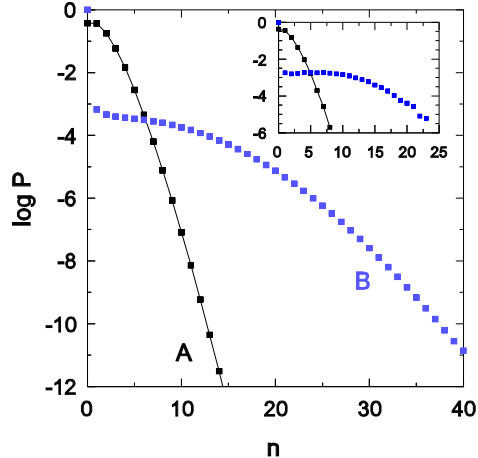}}
\caption{\footnotesize{One-site probability distributions for the number of A
and B particles. Main graph: one-site MFT, $\rho=1$, $r \simeq r_c = 0.688$,
$D_A = 0.5$ and $D_B = 0.02$.  Inset: simulation result for $r=0.5$ and the
same values of $\rho$, $D_A$ and $D_B$, system size $L=500$.}} \label{dist002g}
\end{figure}

The reason for the increase in $r_c$ at small values of $D_B$ appears to lie in
the possibility of accumulating many B particles at a single site.  Suppose
that, by some fluctuation, a site acquires several B particles.  Since $D_B \ll
D_A$, this accumulation is relatively long-lived, and any A particles at this
site will rapidly change to B, since the effective rate for the reaction $A \to
B$, at this site, is $b$, which is large compared to $r$, the rate of the
inverse reaction, and also large compared to $D_A$.  (Note as well that A
particles straying onto such a site will readily change to B, and tend to
remain at this site for a long time.) Indeed, for $D_B < D^*$, mean-field
theory reveals that the one-site marginal distribution $P(b)$, though quite
small for $b \geq 1$ decays very slowly with increasing $b$. The distributions
$P(a)$ and $P(b)$ are compared in Fig.~\ref{dist002g} (for the case $\rho=1$,
$r \simeq r_c = 0.688$, $D_A = 0.5$ and $D_B = 0.02$).  Although (globally)
almost all particles belong to species A, $P(b)$ decays much more slowly than
$P(a)$ for large occupancies.  (In this regime we increase $a_c$ and $b_c$ to
40 in the one-site MF calculations, to ensure that the cutoff does not affect
the result.) $P(a)$ is well approximated by a Poisson distribution, while
$P(b)$ shows strong deviations from this from.  As shown in Fig.~\ref{dist002g}
(inset), similar distributions are observed in simulation, in the regime $D_B
\ll D_A$.

\section{Simulations}

We perform Monte Carlo simulations using a simulation algorithm designed to
reproduce faithfully the transition rates defining the process. Simpler, more
efficient computational models involving the four types of transition [A and B
diffusion, recovery (R), and infection (I)], are possible, but do not
correspond to the same set of transition rates.  Our simulation method permits
quantitative comparison with theoretical predictions, including systematic
expansions of the master equation \cite{inprog}.

The simulation consists of a sequence of events, each of which involves
choosing the type of transition and the site at which it occurs. The choice of
event type depends the total transition rates for each of the four processes,
given by:

\begin{itemize}
\item Hopping of A particles: total transition rate $W_A = N_A D_A$,
where $N_A = \sum_j a_j$ is the total number of A particles.

\item Hopping of B particles: total transition rate $W_B = N_B D_B$.

\item Transformation of B particles to A particles: total transition rate $W_R = r N_B$.

\item Transformation of A particles to B particles: total transition rate
$W_I = \sum_j a_j b_j$.
\end{itemize}

If we let $W_T = W_A + W_B + W_R + W_I$ denote the total transition rate for
all processes, then the probability that the next event is of type $m$ ($=$ A,
B, R or I), is $P_m = W_m/W_T$, while the mean time to the next event is
$\Delta t = 1/W_T$. (After each event the time is advanced by this amount.) The
next event is chosen at random from this set of probabilities.  Once the event
type is determined, the site at which it occurs must be chosen.  For this
purpose a number of lists are maintained. For example, if the chosen event is
hopping of an A particle, we select a site at random from a list of all sites
(call it the A-list) having $a_j > 0$. Not all sites on the A-list, however,
are equally likely to host the event: since each A particle has the same
hopping rate, the probability of the event occurring at site $j$ is
proportional to $a_j$. We therefore keep track of the number $a_{max}$ of A
particles at the site with the largest number of such particles. When site $j$
is chosen from the A-list, it is accepted for the next event with probability
$p_{acc} = a_j/a_{max}$. In case of rejection, we again select a site ($k$,
say) from the A-list and compare $a_k$ with $a_{max}$. In his manner we ensure
that each A particle has the same likelihood of hopping.  In a diffusion event
the target site is chosen at random from the nearest neighbors of the original
site.

The same procedure is adopted for the other three processes, necessitating the
maintenance of a B-list and an AB-list (the latter containing all sites such
that $a_j b_j > 0$).  The rejection procedure outlined above, while necessary
to maintain fidelity to the original transition rates, is computationally
expensive.  For this reason we restrict the present study to a relatively low
density, $\rho =1$, since in this case the maximum values, $a_{max}$,
$b_{max}$, and $(ab)_{max}$ are generally not very large.  (One could in fact
use lists of the positions of all A and B {\it particles}, instead of A- and
B-bearing sites.  But since the A-B reaction step requires an AB site list, we
adopt a uniform procedure for all processes.  Using particle instead of site
lists for all processes except the A-B reaction could improve efficiency,
particularly for large diffusion rates.)

In the studies reported here we sample the {\it quasi-stationary} (QS)
distribution of the process, (that is, conditioned on survival), an approach
that has been found very useful in the study of systems with an absorbing state
\cite{qss}. (In fact, conventional simulations of ``stationary" properties of
models with an absorbing state actually study the quasi-stationary regime,
given that the only true stationary state for a finite system is the absorbing
one). We employ a recently devised simulation method that yields
quasi-stationary properties directly \cite{qssim}. This is done by maintaining,
and gradually updating, a set of configurations visited during the evolution;
when a transition to the absorbing state is imminent the system is instead
placed in one of the saved configurations. Otherwise the evolution is identical
to that of a conventional simulation.  (The set of saved configurations is
updated by replacing one of the saved configurations with the current one, with
a small probability $p_{rep}$ at each time step.)

The above scheme was shown \cite{qssim} to yield precise results, in accord
with the exact QS distribution for the contact process on a complete graph, and
with conventional simulations of the same model on a ring \cite{qssim}. The
scheme has also been shown to yield results that agree, to within uncertainty,
with the corresponding results of conventional simulations for a sandpile model
\cite{qssand}. The advantage of the method is that a realization of the process
can be run to arbitrarily long times. Thus, whereas in conventional simulations
a large number of realizations must be performed to have a decent sampling of
the quasi-stationary state, here every realization provides useful information.
This leads to an order of magnitude improvement in efficiency in the critical
region.  For further details on the method see \cite{qssim}.

We performed extensive simulations of the DEP on rings of $L=200$, 500, 1000,
2000, and 4000 sites, using the QS method. The number of saved configurations
$M$ ranges from 1000, for $L=200$, to $100$ for $L=4000$.  Values of $p_{rep}$
range from $10^{-4}$ to $10^{-5}$  (smaller values for larger systems). Two
time scales appear to be relevant in the choice of the replacement probability
$p_{rep}$.  One is the mean residence time of a configuration on the list,
$\tau_L= M/p_{rep}$. The other is the QS lifetime $\tau$, i.e., the mean time
between attempts to visit the absorbing state.  We find that our results are
independent of the choice of $p_{rep}$ provided $\tau/\tau_L < 1$. This appears
to be associated with the need to preserve configurations visited prior to the
last attempted transition to the absorbing state.  Of course one could make
$\tau_L$ arbitrarily large by reducing $p_{rep}$, but this would prolong the
memory of the initial state.  (For this reason, we use a $p_{rep}$ ten times
larger during the relaxation phase.  The latter represents about 10\% of the
total simulation time.)

Initially, half the particles are of type A and half B; the particles are
distributed randomly and independently over the sites, so that the distribution
of $a$ and $b$ at a given site is essentially Poisson with mean 1/2. Each
realization of the process runs for a certain maximum time, $T_m$, of up to
$10^8$ time units. The results reported here represent averages over 4 - 12
independent realizations for each set of parameter values.  Averages are taken
in the QS regime, after discarding an initial transient, with a duration that
depends on the system size.  In practice we accumulate histograms of the time
during which the system has exactly 1, 2,...n,..., B particles. The histograms
are used to evaluate the mean B particle density, $\rho_B$ and the moment ratio
\cite{dic-jaff}

\begin{equation}
m = \frac{\langle b^2 \rangle}{\langle b \rangle^2} .\label{defm}
\end{equation}

We determine the critical recovery rate $r_c(\rho,D_A,D_B)$ using the criteria
of power-law dependence of $\rho_B$ and $\tau$ on system size $L$ (i.e., the
usual finite-size scaling relations $\rho_B \sim L^{-\beta/\nu_\perp}$ and
$\tau\sim L^z$). To probe the three characteristic regimes of the process,
three cases are studied in detail: $D_A = D_B = 0.5$; $D_A = 0.5$, $D_B=0.25$;
and $D_A = 0.25$, $D_B = 0.5$.

\begin{figure}[h]
\epsfysize=8cm \epsfxsize=10cm \centerline{ \epsfbox{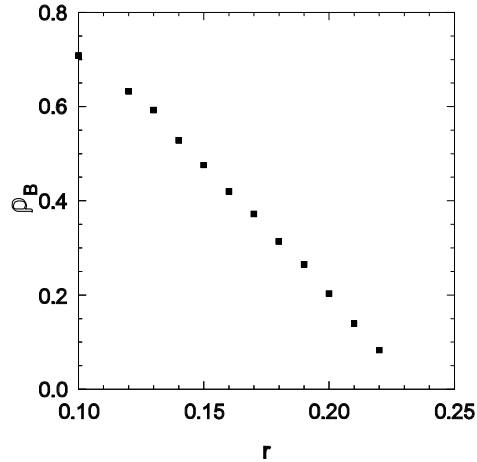}}
\caption{\footnotesize{Order parameter $\rho_B$ versus recovery rate for
$D_A=0.5$, $D_B=0.25$, and $\rho=1$.  The points represent extrapolations to
the infinite-size limit based on data for sizes $L=200$ - 4000.}}
\label{epd525}
\end{figure}

Consider first the case $D_A = 0.5$, $D_B = 0.25$.  Since it has been suggested
that the phase transition is discontinuous for $D_A > D_B$, it is of interest
to study the behavior of the order parameter $\rho_B$ as a function of the
recovery rate $r$.  In Fig. \ref{epd525} we plot estimates for the order
parameter in the limit $L \to \infty$, based on our data for $L$ = 200 - 4000.
(The extrapolation can only be performed reliably for $r \leq 0.22$ with the
data at hand.)  It appears that the order parameter decays {\it continuously}
to zero as $r \to r_c \simeq 0.23$.  (We cannot rule out a weakly discontinuous
transition on the basis of these data, but a continuous transition seems the
more natural interpretation.)  The evidence for a continuous transition is
greatly strengthened by our observation of critical scaling, as we now discuss.

\begin{figure}[h]
\epsfysize=8cm \epsfxsize=10cm \centerline{ \epsfbox{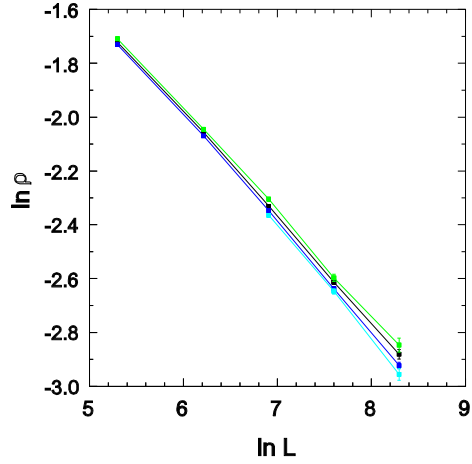}}
\caption{\footnotesize{Order parameter $\rho_B$ versus system size $L$, for the
same parameters as Fig. 4.  From top to bottom: recovery rate $r= 0.231$,
0.232, 0.233, 0.234.}} \label{fssrho}
\end{figure}

\begin{figure}[h]
\epsfysize=8cm \epsfxsize=10cm \centerline{ \epsfbox{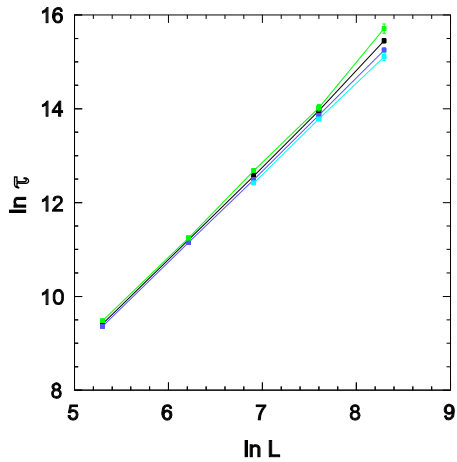}}
\caption{\footnotesize{Lifetime $\tau$ versus system size $L$, symbols and
parameters as in Fig. 5.}} \label{fsstau}
\end{figure}

Data for the QS order parameter $\rho_B$ versus system size $L$ are shown on
log scales (for the case $D_A = 0.5$, $D_B=0.25$) in Fig.~\ref{fssrho}.  The
data for $r=0.231$ curve upward, while those for $r=0.234$ curve downward,
leading to the estimate $r_c = 0.2325(10)$.  For this range of values we see
good evidence of power-law scaling, as expected at a critical point. Similarly,
the data for the QS lifetime $\tau$ (see Fig.~\ref{fsstau}) show power-law
scaling for $r \simeq r_c$, and significant curvature away from $r_c$.

As a check on our procedure for determining $r_c$, we also perform, for $D_A =
0.5$, $D_B=0.25$, initial decay studies \cite{kockelkoren,parkpark06}.  In
these studies the order parameter $\rho_B$ is followed as a function of time,
using the same initial condition as in the QS studies. (Due to the large system
size, the process does not enter the absorbing state on the time scale of the
simulation.) Here the expected critical behavior is $\rho_B \sim t^{-\theta}$.
Using deviations from the power law to identify off-critical values, a study of
systems of $10^6$ sites, to a maximum time of 10$^7$, enables us to restrict
$r_c$ to the interval [0.231, 0.239].  This is consistent with, but less
precise than, the QS results. Analysis of the data for $t \geq 5 \times 10^5$
furnishes $\theta \simeq 0.5$. The lack of precision in these studies and in
our result for the moment ratio highlights the slow convergence of these
simulations, presumably reflecting strong finite-size effects.

The studies described above lead to estimates for the critical exponent ratios
$\beta/\nu_\perp$ and $z = \nu_{||}/\nu_\perp$.  To obtain a direct estimate of
$\nu_\perp$, we study the order parameter in the neighborhood of the critical
point $r_c$ for various lattice sizes. This permits us to estimate the
correlation length exponent $\nu_\perp$, using the finite-size scaling relation
for the order parameter,

\begin{equation}
\rho_B(\Delta,L)\propto L^{-\beta/\nu_\perp}\mathcal{F}
(L^{1/\nu_\perp}\Delta),
\end{equation}
where $\Delta = r-r_c$ and $\mathcal{F}$ is a scaling function. This implies
\begin{equation}
\left| \frac{\partial \ln \rho_B}{\partial r}\right|_{r_c} \propto
L^{1/\nu_\perp}.
\end{equation}

\noindent For the case $D_A = 0.5$, $D_B=0.25$, the data for $L=$ 500, 1000,
2000 and 4000 yield the estimate $\nu_\perp= 2.3(3)$. We performed similar
finite-size scaling analyses for the other cases.

\begin{figure}[h]
\epsfysize=8cm \epsfxsize=10cm \centerline{ \epsfbox{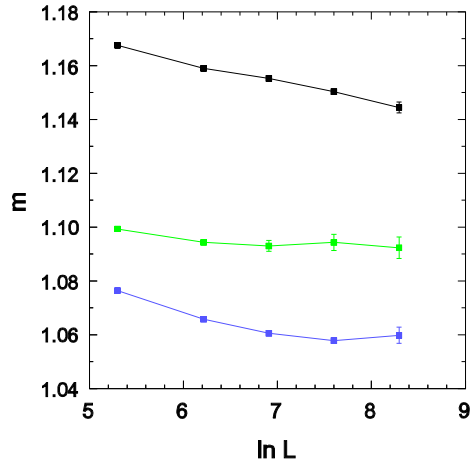}}
\caption{\footnotesize{Moment ratio $m$ versus system size $L$, at the critical
point $r_c$, for (upper to lower) $D_A = 0.5$, $D_B=0.25$; $D_A = D_B = 0.5$;
and $D_A=0.25$, $D_B=0.5$.}} \label{fssm}
\end{figure}

The moment ratio $m$ is also useful in characterizing critical behavior. Our
results, shown in Fig.~\ref{fssm}, lead to estimates for the limiting ($L \to
\infty$) value except in the case $D_A = 0.5$, $D_B =0.25$, for which the
moment ratio decreases systematically with system size $L$.  The trend to
smaller values of $m$ with increasing $D_B/D_A$ may reflect the reduced
tendency for B particles to accumulate at a given site. Our results for the
critical parameters are summarized in Table II.

We close this section with the observation that the DEP is characterized by
unusually large and long-lived fluctuations, which make it difficult to obtain
precise results in simulations.  In the case of the contact process, for
example, a system size of $L=1000$ is sufficient for the determination of
critical exponents to rather high precision.  As in other models dominated by
diffusion, such as conserved sandpiles \cite{qssand} and the diffusive pair
contact process (PCPD) \cite{henkelhinrichsen,pcpdqss}, scaling properties of
the DEP emerge clearly only at larger sizes.  One signal of this is the slow
convergence (with $L$) of the order parameter moment ratio $m$.

\section{Discussion}

We study the one-dimensional diffusive epidemic process using mean-field
approximations and Monte Carlo simulations.  Mean-field theory provides a
qualitative description of the phase diagram, and reveals a surprising,
non-monotonic dependence of the critical recovery rate on the B particle
diffusion rate, confirmed in simulation.  This reflects the tendency (for $D_B
\ll D_A$) for many B particles to accumulate at a single site, as revealed in
the one-site probability distribution $P(b)$. Mean-field theory also captures
qualitatively the anti-correlation in the number of A and B particles at a
single site.

Although our simulations do not extend to sufficiently large systems to furnish
precise values of critical exponents, our results, especially for the ratio
$\beta/\nu_\perp$, clearly support the scenario of distinct universality
classes for the three cases, $D_A < D_B$, $D_A = D_B$, and $D_A > D_B$.  The
transition appears to be continuous in all cases.  Renormalization group
analysis \cite{kree,wijland,oerding,janssende} predicts (independent of the
relative magnitudes of $D_A$ and $D_B$) that $z=2$ and $\nu_\perp = 2/d$.  The
simulation results are consistent with these predictions {\it except} for the
case $D_A = 0.25$, $D_B = 0.5$, for which our estimates are significantly
smaller.  This may be due to the slow convergence noted above.  We hope to
address this point in future studies of larger systems.  It would also be of
interest to verify that the critical exponents are in fact insensitive to
changes in the diffusion rates, within the three universality regimes that have
been established. We expect the present results to serve as a point of
reference for studies based on systematic expansions of the master equation.

\vspace{1em}

\noindent{\bf Acknowledgments}

We are grateful to Hugues Chat\'e for helpful comments. This work was supported
by CNPq and Fapemig, Brazil.

\newpage

\bibliographystyle{apsrev}

\newpage

\begin{center}
{\sf Table I. Critical recovery rate $r_c$ in one- and two-site\\ mean-field
approximations, compared with simulation.} \vspace{1em}

\begin{tabular}{c c c c c}
\hline \hline
 $D_A$ & $D_B$ & $\;\;\;\; r_c$ (1-site) & $\;\;\;\; r_c$ (2-site) & $\;\;\;\; r_c$ (sim) \\
\hline
 0.5   & 0.25  &  0.5420        & 0.411          & 0.2325(10)  \\

 0.5   & 0.5   &  0.5771        & 0.429          & 0.1921(5)  \\

 0.25  & 0.5   &  0.5144        & 0.368          & 0.1585(3)   \\
\hline \hline
\end{tabular}
\end{center}
\vspace{3em}

\begin{center}
{\sf Table II. Critical parameters from simulation.} \vspace{1em}

\begin{tabular}{c c c c c c}
\hline \hline
$ D_A$ & $D_B$   &  $\beta/\nu_\perp$ & $z$ & $\nu_\perp$ & $m$\\
\hline

0.5  & 0.25 &   0.404(10) & 2.01(4)  & 2.3(3) & $<$ 1.15\\

0.5  & 0.5  &   0.192(4)  & 2.02(4)  & 2.0(2) & 1.093(10)\\

0.25 & 0.5  &   0.113(8)  & 1.6(2)   & 1.77(3)& 1.06(1)\\

\hline \hline
\end{tabular}
\end{center}

\end{document}